\documentclass[aps,pra,letterpaper,10pt, twocolumn,showpacs,superscriptaddress]{revtex4-1}
\usepackage{hyperref}
\usepackage{amssymb}
\usepackage{amsmath}
\usepackage{graphicx}
\usepackage{subfigure}
\usepackage{amsfonts}
\usepackage{xcolor}
\usepackage{epsfig}
\usepackage{color}
\usepackage{bbm}

\newcommand{\ket}[1]{\vert{#1}\rangle} 
\newcommand{\bra}[1]{\langle{#1}\vert} 
 
\newcommand{\proj}[1]{\ket{#1}\!\bra{#1}}
\newcommand{\trans}[2]{\ket{#1}\!\bra{#2}}
\newcommand{\mean}[1]{\langle #1 \rangle}

\newcommand{\abs}[1]{\left|#1\right|}
\newcommand{\sighat}{\hat{\sigma}}

\newcommand{\xzpm}{x_{\rm zpm}}
\newcommand{\Nbos}{\overline{N}}
\newcommand{\rhotd}{\mu_{2\rm D}}
\newcommand{\tr}[1]{\mathrm{Tr}#1}

\newcommand{\liov}{\mathcal{L}}
\newcommand{\diss}{\mathcal{D}}

\newcommand{\hatd}[1]{\hat{#1}^{\dagger}}

\begin{document}
\title{Dissipative optomechanical preparation of macroscopic quantum superposition states}

\author{M. Abdi}
\affiliation{Physik Department, Technische Universit{\"a}t M{\"u}nchen, James-Franck-Str.~1, 85748 Garching, Germany}
\affiliation{Institut f{\"u}r Theoretische Physik, Albert-Einstein-Allee 11, Universit{\"a}t Ulm, 89069 Ulm, Germany}

\author{P. Degenfeld-Schonburg}
\affiliation{Physik Department, Technische Universit{\"a}t M{\"u}nchen, James-Franck-Str.~1, 85748 Garching, Germany}

\author{M. Sameti}
\affiliation{Institute of Photonics and Quantum Sciences, Heriot-Watt University, Edinburgh, EH14 4AS, United Kingdom}

\author{C. Navarrete-Benlloch}
\affiliation{Max-Planck-Institut f{\"u}r Quantenoptik, Hans-Kopfermann-Str.~1, 85748 Garching, Germany}
\affiliation{Institute for Theoretical Physics, Universit{\"a}t Erlangen-N{\"u}rnberg, Staudtstrasse 7, 91058 Erlangen, Germany}

\author{M. J. Hartmann}
\affiliation{Institute of Photonics and Quantum Sciences, Heriot-Watt University, Edinburgh, EH14 4AS, United Kingdom}

\date{\today}
\begin{abstract}
The transition from quantum to classical physics remains an intensely debated question even though it has been investigated for more than a century. Further clarifications could be obtained by preparing macroscopic objects in spatial quantum superpositions and proposals for generating such states for nano-mechanical devices either in a transient or a probabilistic fashion have been put forward. Here we introduce a method to deterministically obtain spatial superpositions of arbitrary lifetime via dissipative state preparation. In our approach, we engineer a double-well potential for the motion of the mechanical element and drive it towards the ground state, which shows the desired spatial superposition, via optomechanical sideband cooling. We propose a specific implementation based on a superconducting circuit coupled to the mechanical motion of a lithium-decorated monolayer graphene sheet, introduce a method to verify the mechanical state by coupling it to a superconducting qubit, and discuss its prospects for testing collapse models for the quantum to classical transition.
\end{abstract}

\pacs{85.85.+j, 03.65.Ud, 42.50.Wk, 85.25.-j}
\maketitle

%
%
A fundamental open question in modern quantum mechanics is how classical physics arises from it as one moves from the microscopic to the macroscopic world. \textit{Decoherence} is arguably the strongest candidate for the process inducing such a transition \cite{Zurek1991}. There are however theories which explain this transition via so-called \textit{collapse} or \textit{spontaneous reduction} models and attribute it to other sources, e.g. spontaneous localization \cite{Bassi03}, quantum \cite{QG2} or classical \cite{CG1} gravity, or uncertainty relations on the space-time continuum \cite{URST}. Proposals for testing such models exist \cite{Arndt2014}, and most of them require the preparation of massive objects either in superposition \cite{Bose1999,Marshall2003,Schwab2005} or entangled states \cite{Hartmann2008,Asadian2015} (the latter also have applications for quantum information processing \cite{Savelev2006,Rips2012,Rips2013}). Several works have emerged proposing protocols for the preparation of such states, especially in the context of optomechanics, either probabilistically \cite{ORIprl11,ORIprl12,Asadian2015} or in the transient regime \cite{Werner2003,Voje2012,Buchmann2012,Ghobadi2014,Abdi2015,Liao2015}. Mechanical elements, however, are exposed to high decoherence rates induced by their finite temperature environments, which demand fast, yet accurate, state preparation and certification methods \cite{Abdi2012}. With direct full-state tomography being quite challenging, some indirect reconstruction methods have been proposed \cite{Clerk2008,Singh2010,Vanner2011}.

Another way to meet the challenge of large mechanical decoherence is to generate the desired states dissipatively, that is, as robust and long-lived steady-states. Here we propose a method for the dissipative preparation of a mechanical element in a superposition of two spatially separated states, together with an efficient way to verify such preparation. Our proposal, therefore, paves the way towards experimental tests of collapse models for the quantum to classical transition. 

We here consider state-of-the-art superconducting circuits and electromechanical devices to show that a highly controllable and tuneable double-well potential can be engineered electrostatically, while the mechanical motion of the trapped element can be cooled to its ground state with high fidelity. Due to the shape of the potential, this ground state is a spatial superposition state. We show how to verify its preparation via population measurements on a single qubit and discuss the avenues this opens for testing collapse models for the quantum to classical transition.

%
%
\textit{Model.---}
We consider the motion of a mechanical element with effective mass $m$ moving in a symmetric double-well potential as described in terms of its position $\hat x$ and momentum $\hat p$ by the Hamiltonian
\begin{equation}
\hat H_{\rm dw} = \frac{\hat p^2}{2m} - \frac{\nu}{2} \hat x^2 +\frac{\beta}{4} \hat x^4.
\label{dw}
\end{equation}
Here, the double-well potential results from a combination of an inverted parabola generating a potential barrier at the origin and an attractive quartic potential that dominates at large deflections. We describe the physical origin of the parameters $\nu$ and $\beta$ for our proposed implementation below and denote the eigenstates and eigenvalues of $\hat H_{\rm dw}$ by $\hat H_{\rm dw}\ket{n}=E_n\ket{n}$ ($n=0,1,2,\dots$).

\begin{figure}[t]
\includegraphics[width=\columnwidth]{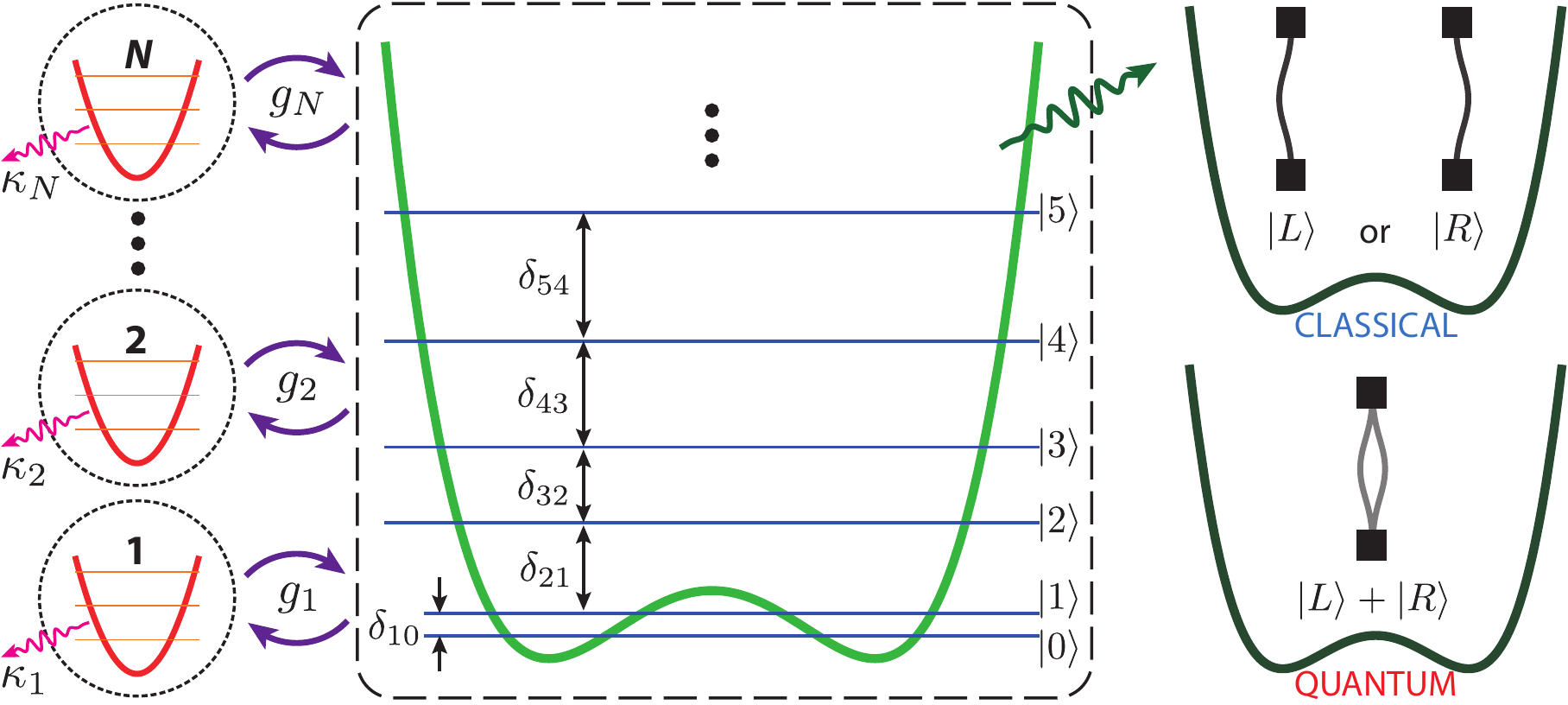}
\caption{(color online) Conceptual scheme of our proposal. A mechanical element moving in a shallow double-well potential with only two possible states below the central barrier is coupled to several electromagnetic modes capable of cooling its different transitions. Whereas the ground state of this potential is a spatial superposition of
left and right deflections, $\ket{0} = (\ket{L}+\ket{R})/\sqrt{2}$, without sideband cooling on several transitions only the classical mixture $(\proj{L}+\proj{R})/2$ would be accessible.}
\label{fig:model}
\end{figure}

The potential of Eq.~(\ref{dw}) together with its eigenvalues is sketched in Fig.~\ref{fig:model}. For states with energies below the peak of the central barrier, tunneling between the wells breaks the left/right degeneracy, and the eigenstates of the system are formed by symmetric and anti-symmetric superpositions of states localized in the two individual wells. These localized states are well approximated by eigenstates of harmonic potentials with frequency $\omega_0=\sqrt{2\nu/m}$ and minima at the well positions $ \pm x_0= \pm \sqrt{\nu/\beta}$. The two lowest energy eigenstates of the double-well potential are thus even and odd cat states, respectively. Preparing the system in either of them results in a quantum superposition of two macroscopically distinguishable states of a massive object, provided that $x_0$ exceeds the zero-point motion $x_\text{zpm}=\sqrt{\hbar/2\omega_0 m}$ associated to the ground state of each well \cite{Monroe1996}.

We now show how optomechanical sideband cooling can be used for ground state preparation in such a highly nonlinear potential. Since the coupling to a thermal environment ensures that the populations of the double-well eigenstates decay exponentially with the energy, it would in principle be sufficient to transfer all population of the first excited state to the ground state. In practice, however, it is challenging to generate a sufficient cooling rate for the lowest transition. The (linearized) coupling between the mechanical oscillator and a cavity electromagnetic mode with annihilation operator $\hat a$ and frequency $\omega_\text{c}$ takes the form $\hbar g\hat{x}(\hat a +\hatd a)$, where $g=g_0\sqrt{\bar{n}_\text{c}}$, with $g_0=(\partial\omega_\text{c}/\partial x)_{x=0}$ the bare optomechanical coupling and $\bar{n}_\text{c}$ the photon number induced in the cavity by an external field driving it at frequency $\omega_c|_{x=0}+\Delta$ that allows to control $g$ \cite{Wilson-Rae2007,Marquardt2007}. Efficient ground state cooling requires essentially three conditions: (\textit{i}) the detuning $\Delta$ of the external driving field with respect to the cavity mode has to be chosen as $\Delta\approx -\delta_{10} \equiv (E_{1}-E_{0})/\hbar$, with (\textit{ii}) a photon relaxation rate $\kappa$ satisfying $\delta_{10}\gg\kappa\gg g_{10}$, while (\textit{iii}) keeping the cooperativity $g_{10}^2/\kappa\gamma_{10}\bar{N}(\delta_{10})$ large enough (here $g_{10}=g\bra 1\hat x\ket 0$ is the optomechanical coupling rate for the lowest mechanical transition, $\gamma_{10}$ is the relaxation rate of the transition, and $\bar{N}(\Omega)$ are the thermal environmental excitations at the corresponding frequency, see below). Yet, as $\delta_{10}$ decreases exponentially with the separation between the wells, the cooperativities and cooling rates generated in this way would be rather limited.

To obtain more efficient cooling, the mechanical system can be coupled to a set of cavity modes with relaxation rates $\kappa_j$, each performing sideband cooling in one or more transitions as illustrated in Fig.~\ref{fig:model}. As we show numerically, this arrangement cools down the mechanical mode close to its ground state even for moderate relaxation rates $\kappa_j\lesssim\delta_{mn}$, where $\delta_{mn}=(E_m -E_n)/\hbar > 0$ refers to the transition we intend to cool with mode $j$. For realizations where all employed cavity modes have comparable linewidths $\kappa_{j}$, as is the case in most settings, the performance of this cooling concept improves with the number of modes. This may however face practical limitations (see below). We remark that our approach does not only apply to model (\ref{dw}) but to any nonlinear potential with a non-degenerate ground state.

In the following we will describe the system via a master equation for its state $\hat{\rho}$. Whereas the dissipation of photons can be treated in the common way, the nonlinearity of the mechanical motion requires special attention. As a master equation is based on a perturbative expansion in system-environment couplings, it here leads to non-unitary terms of Lindblad form for each transition between eigenstates of the mechanical mode \cite{Breuer2007,Beaudoin2011,Ridolfo12,Ridolfo13,supmat}, since each transition couples with a different strength to the environment and experiences a different density of states. Using an adequate `microscopic' model for the system-environment interaction, we thus derive the master equation \cite{supmat},
\begin{equation}
	\partial_t\hat{\rho} = \frac{1}{i\hbar}\left[\hat H,\hat{\rho}\right] +\frac{\kappa}{2}\sum_j\mathcal D_{\hat a_j}[\hat{\rho}] +\frac{1}{2}\liov_{\rm m}[\hat{\rho}],
\label{master}
\end{equation}
where $\hat H = \hat H_{\rm dw} +\sum_j[-\hbar\Delta_j\hatd a_j\hat a_j +\hbar g_j(\hat a_j +\hatd a_j)\hat x]$ is the full Hamiltonian and $\mathcal D_{\hat O}[\cdot] = 2\hat O (\cdot) \hatd O -\hatd O\hat O(\cdot) -(\cdot)\hatd O\hat O$ a standard Lindblad superoperator. $\liov_m[\cdot]=[\hat x, (\cdot)\hatd A -\hat A(\cdot)]$ is the mechanical dissipator \cite{supmat} with
\begin{equation*}
	\hat A = \sum_{m>n}\gamma_{mn} k_{mn}\Big\{\bar{N}(\delta_{mn})\trans{m}{n} +[\bar{N}(\delta_{mn})+1]\trans{n}{m}\Big\},
\end{equation*}
where $\gamma_{mn}=\delta_{mn}/Q$ is the $\ket{m} \rightarrow \ket{n}$ decay rate and $\bar{N}(\Omega)=[\exp(\hbar\Omega/k_{\rm B}T)-1]^{-1}$ the reservoir occupation at temperature $T$ and frequency $\Omega$. $Q$ is the quality factor of harmonic mechanical oscillations and $k_{mn} = x_{mn}(2m\omega/\hbar)$ are the position matrix elements normalized to the zero-point position variance associated to the original harmonic oscillations, see below.
We now turn to propose a specific implementation of these ideas, for which a realistic choice of parameters allows
to cool the mechanical mode to the ground state showing the desired superposition.

\begin{figure}[t]
\includegraphics[width=\columnwidth]{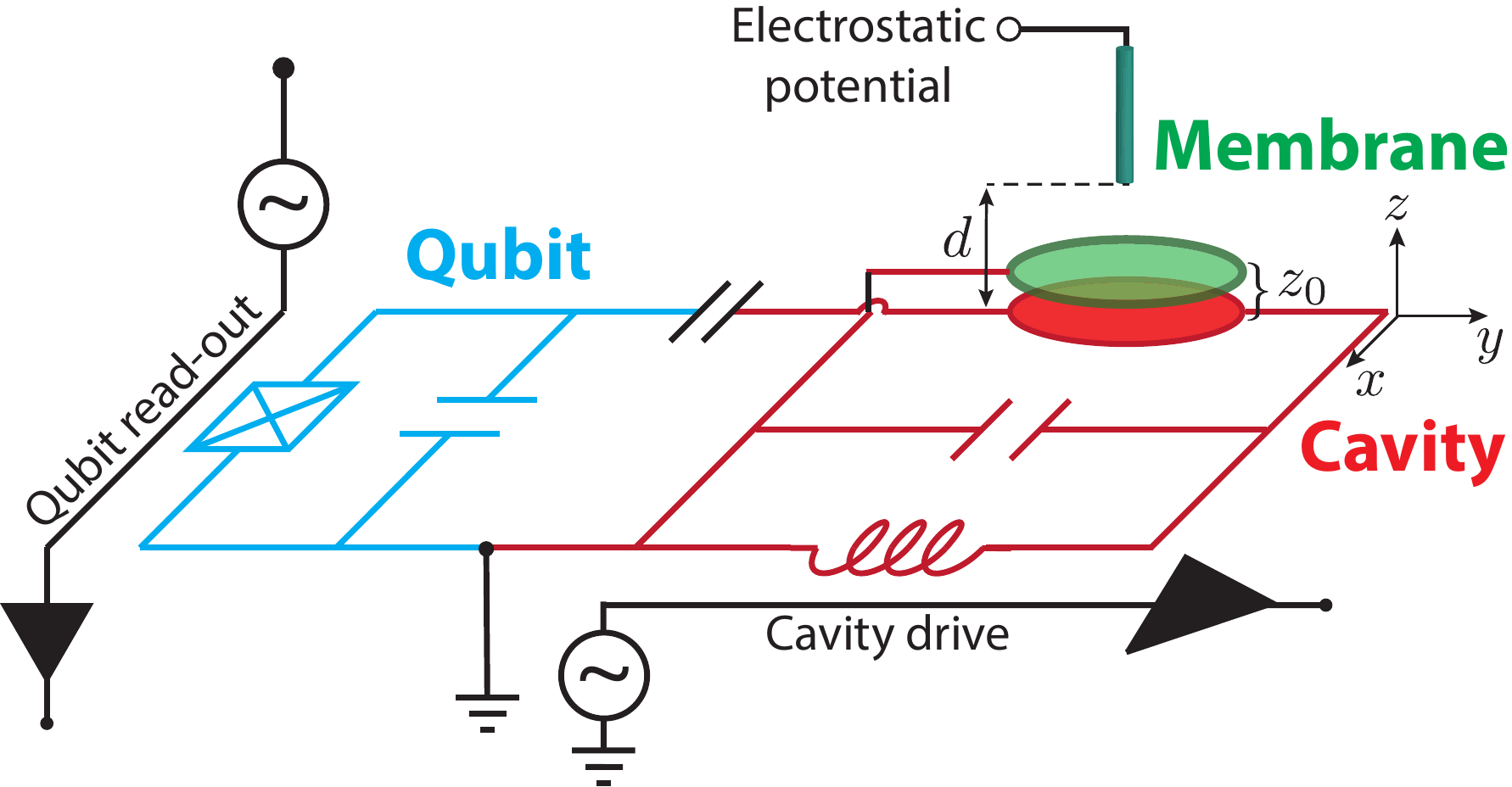}
\caption{(color online) Scheme for the circuit quantum electromechanical implementation of our proposal. The mechanical membrane (green) alongside a disk electrode of a microwave cavity (red) produces a position-dependent capacitance, and hence an optomechanical coupling to the cavity modes. A superconducting qubit (blue) is capacitively coupled to the cavity, and we employ it for reading out the mechanical state. The double-well potential is engineered on the membrane through an electrostatically fed electrode tip placed above its center, which produces a softening potential.}
\label{fig:scheme}
\end{figure}

%
%
\textit{Implementation.---}
To implement our ideas, we here propose an architecture based on state-of-the-art superconducting circuits and electromechanical technology, see Fig.~\ref{fig:scheme} and Ref.~\cite{Lecocq2015} for a similar device. In this proposal, the mechanical degree of freedom is realized by the drum mode of a thin circularly clamped mechanical layer (membrane) of radius $a$, which is confined in a double-well potential generated via the electrostatic field of a tip electrode located above its center. The cavity modes, in turn, are the resonance modes of a superconducting resonator with a disk-shaped end placed below the membrane such that plate and membrane form a capacitance. The dependence of its capacitance on the plate-to-membrane separation then provides the desired optomechanical interaction. An additional superconducting qubit coupled to the microwave resonator will allow us to read out the mechanical motion.

Let us now elaborate on the physics behind this scenario. In \cite{supmat} we provide a detailed derivation of the Hamiltonian describing the membrane. The fundamental flexural mode with position $\hat x$ and momentum $\hat p$ is found to be well isolated from the rest of the mechanical modes and, besides a harmonic restoring potential with frequency $\omega$, it is subject to a nonlinear potential of geometric origin that is, to leading order, well-described by a Duffing nonlinearity of the form $\beta\hat x^4/4$. The total elastic potential $\hat V_{\rm m} =m \omega^{2} \hat x^{2}/2 + \beta\hat x^4/4$ can be transformed into a double-well of the form (\ref{dw}) by adding a `softening' force which generates an additional potential with the shape of an inverted parabola that exceeds the harmonic confinement in $\hat V_{\rm m}$.
A feasible way to controllably generate such a force on the membrane consists in applying an inhomogeneous electrostatic field generated via a tip electrode situated close to its center, see Fig.~\ref{fig:scheme}. The electrostatic energy of the membrane's fundamental mode can be expanded in its deflection, $\hat V_{\rm es} = \sum_{j=1}^\infty \alpha_j \hat x^j$, with coefficients $\alpha_j = \pi h \epsilon_0 \int_0^a rdr \partial^j_z E_z^2(r,z=z_0)\psi_0^j(r)$ \cite{supmat}. Here, $\psi_0$ is the profile of the fundamental mode, $E_{z}$ is the static electric-field component perpendicular to the mechanical layer, and $h$ is the membrane's thickness at rest in the $z=z_0$ plane ($z=0$ is taken at the superconducting disk below the membrane).
$\alpha_1$ shifts the equilibrium position of the mechanical mode and can be used as an additional control knob for the coupling to the electromagnetic modes \cite{Weber2014}. $\alpha_2$ can be made negative, therefore generating the softening force that leads to the double-well potential. Higher orders are shown to be negligible \cite{supmat}.
Hence, we see that the combination of the geometrical and electrostatic potentials, $\hat V_{\rm m} +\hat V_{\rm es}$, gives rise to the desired double-well potential for the membrane's motion, with a parameter $\nu = 2|\alpha_2| -m\omega^2$ that can be tuned via the applied electrostatic fields.

As ideally suited candidates for the mechanical elements we here consider \textit{monolayer graphene} sheets, since they have mechanical properties adapted to our needs. In particular, their ultra-low mass provides them with large zero-point motion $x_\text{zpm}$, while their large Young modulus confers them large Duffing nonlinearity $\beta$. Such sheets have already been studied as mechanical resonators \cite{Weber2014,Singh2014,Song2014}. Importantly, we note that the low conductivity of graphene that has limited optomechanical cooling in these experiments can be overcome by doping the membrane surface with alkaline-metal atoms, as has been recently shown with lithium-decorated graphene (LDG) sheets \cite{Ludbrook2015}, which possess almost the same mechanical properties as standard monolayer graphene \cite{Qi2010} but are superconducting \cite{Profeta2012}.

We now turn to discuss the achievable fidelity for preparing stationary spatial superposition states with our approach for parameters corresponding to this specific implementation. The superconducting gap imposes a limit to the number of high quality resonance modes in the cavity. We thus consider three cavity modes coupled to three \textit{proper} mechanical transitions. Assuming the same decay rate $\kappa = \kappa_{j}$ ($j = 1,2,3$) for all these modes, we tune the optomechanical couplings $g_j$ such that the probability of being in the ground state of the double-well is maximized.

%
%
\textit{Results.---}
We consider a monolayer LDG sheet with radius $a=1~\mu\text{m}$, and thus $m\approx 5.7\times 10^{-16}$~gr, $\omega/2\pi \approx 26$~MHz, and $\beta \approx 5.7\times 10^{15}$~J/m$^4$ \cite{supmat,Singh2014}. The distance between the membrane and the disk-shaped end of the cavity is taken to be $z_0=100$~nm. For a cavity with fundamental resonance at $5$~GHz we get the `bare' optomechanical coupling rate $G_0 = g_0\sqrt{\hbar/2m\omega} \approx 2\pi \times 10$~Hz. As $G_0$ scales linearly with the diameter of the membrane \cite{Weber2014}, its small radius should be compensated by an increased intracavity photon number $\bar{n}_\text{c}$.
The electrostatic field is adjusted such that $\alpha_2 = -1.000134 (m\omega^2/2)$, which requires the application of a few hundred volts to an antenna of a few hundred nanometers size (similar to the tip of a scanning tunneling microscope) located about 1~$\mu$m above the center of the membrane, and creates a shallow double-well with only two levels below the barrier, c.f. Fig.~\ref{fig:model}(a). This situation assures a reasonably large value of $\delta_{10}$ ($\sim 2\pi\times 50$~kHz) that allows us to access the required resolved sideband regime.

\begin{figure}[t]
\includegraphics[width=\columnwidth]{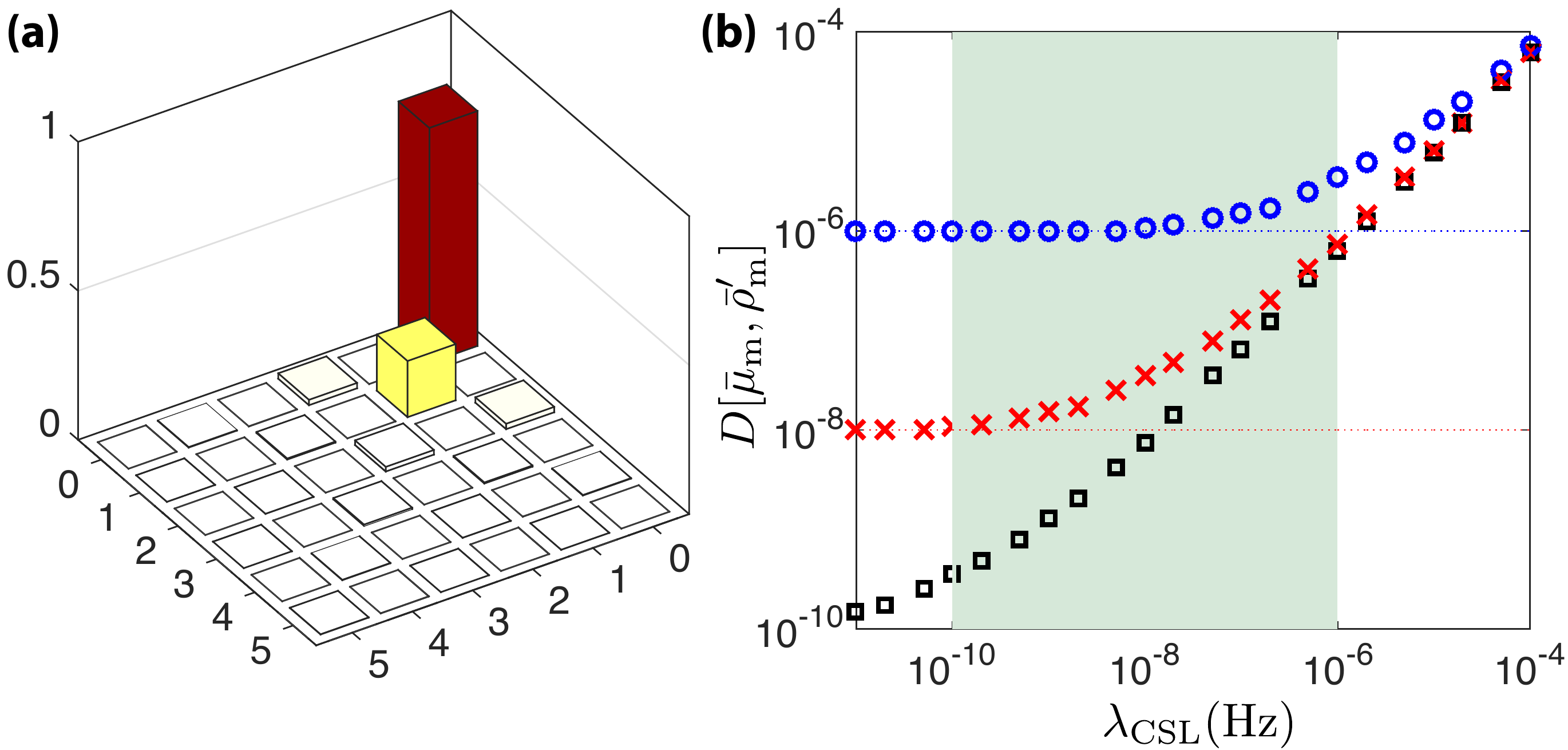}
\caption{(color online) (a) Absolute value of the lowest elements of the mechanical steady-state density matrix when cooled by three electromagnetic modes with optimal detunings and optomechanical couplings as explained in the text. (b) Distance between the population distributions of steady states $\bar\rho_{\rm m}'$ and $\bar\mu_{\rm m}$ in the absence or presence of CSL versus localization rate $\lambda_{\rm CSL}$ for measurement precision $\sigma = 10^{-6}$ (circles), $10^{-8}$ (crosses), and $10^{-10}$ (squares). The shaded area is the predicted range for $\lambda_{\rm CSL}$ \cite{Bassi03,Bahrami2014,Vinante2015}.}
\label{fig:results}
\end{figure}

We numerically \cite{NavarreteNumerics15} obtain the steady-state solution $\bar{\rho}$ of the master equation (\ref{master}), and analyze the ground state population $P_{00}=\langle 0|\bar{\rho}_\mathrm{m}|0\rangle$ present in the reduced mechanical state $\bar{\rho}_\mathrm{m}=\tr_\text{cavity}\{\bar{\rho}\}$.
We consider three cavity modes with decay rates $\kappa = 0.3\delta_{10}$, detunings matching the mechanical transitions $\ket{1} \leftrightarrow \ket{0}$, $\ket{3} \leftrightarrow \ket{0}$, and $\ket{2} \leftrightarrow \ket{1}$,  (i.e. $\Delta_1=-\delta_{10}$, $\Delta_2=-\delta_{30}$, and $\Delta_3=-\delta_{21}$), and optimized intra-cavity photon numbers $\bar{n}_{\text{c}1} = 1200$, $\bar{n}_{\text{c}2} = 1100$, and $\bar{n}_{\text{c}3} = 4000$ \cite{supmat}. Note that only transitions with $x_{mn} \neq 0$ can be cooled which requires $m$ and $n$ to have different parity \cite{supmat}. Assuming an environment temperature of $T=15$ mK and a quality factor $Q=10^6$ of the membrane in the original approximately harmonic potential $\hat{V}_\text{m}$, we obtain $P_{00}\approx 0.79$, meaning that the mechanical resonator can be found in the desired spatial-superposition state with $\sim79\%$ probability [Fig.~\ref{fig:results}(a)].
This non-equilibrium steady-state of the mechanical mode has a spatial extent equal to the separation of the wells, whose ratio to the zero-point motion amplitude in each well is $2 x_0/x_\text{zpm}\approx 6$. It is reached in about 30~$\mu$s, which is orders of magnitude shorter than the time scales of fluctuations in the electrostatic control fields or microwave tones, see Ref.~\cite{Rips2014}. Larger probabilities can be obtained by working deeper in the resolved sideband regime and/or by employing more cavity modes.


To show how our proposal could be exploited for the examination of unconventional sources of decoherence, let us consider the bounds it imposes on the continuous spontaneous localization (CSL) model \cite{Bassi03,Bahrami2014,Vinante2015} as the most prominent collapse model. The CSL model is characterized by a localization length usually taken to be $r_{\rm CSL}=$100~nm, and a localization rate which is predicted to be in the $\lambda_{\rm CSL} = 10^{-8 \pm 2}$~Hz range \cite{Bassi03}. In the limit where the delocalization amplitude $x_0$ is much smaller than $r_{\rm CSL}$, the effect of the CSL model can be approximated by a momentum diffusion term (CSL diffusion) of the form $-(\lambda_{\rm CSL}\eta/r_{\rm CSL}^2)\big[\hat x,[\hat x,\hat\rho]\big]$ ($\eta\approx 1.2\times 10^{15}$ is a mass and geometry dependent factor \cite{Nimmrichter2014}) to be added to equation (\ref{master}). 

Our setup can be used to distinguish CSL diffusion from conventional sources of noise by revealing the differences between the steady-states in the presence and absence of CSL. To this end, one would proceed as follows. After a sufficient experimental characterization of the setup, i.e. its mechanical spectrum, optomechanical couplings etc., one can infer the mechanical damping in a sideband cooling experiment, see e.g. \cite{Jockel11,Groblacher:2015fk}. Whether only the assumed mechanical damping or possibly also CSL is present in the experiment can then be determined by measuring several matrix elements of the mechanical steady-state, see below for a measurement method. To quantitatively analyze this procedure we calculated the steady-state $\bar\mu_{\rm m}$ for a mechanical quality factor $Q$ in the presence of CSL and the steady-state $\bar\rho_{\rm m}'$ in the absence of CSL but with a mechanical quality factor $Q' < Q$, chosen such that the ground state occupations are equal $\bra 0 \bar\rho_{\rm m}'\ket 0 = \bra 0 \bar\mu_{\rm m}\ket 0$. To mimic a finite measurement precision $\sigma$, we only require that $|\bra 0 \bar\rho_{\rm m}'\ket 0 - \bra 0 \bar\mu_{\rm m}\ket 0 | \lesssim \sigma$. Crucially, the CSL diffusion rate $\lambda_{\rm CSL}\eta/r_{\rm CSL}^2$ is independent of the mechanical spectrum, whereas the thermal damping rates $\gamma_{mn}$ strongly depend on the anharmonicity of the potential \cite{supmat}. As a consequence, the occupation probabilities of excited mechanical states differ, i.e. $| \bra j \bar\rho_{\rm m}'\ket j - \bra j \bar\mu_{\rm m}\ket j | \gg \sigma$ for $j \ge 1$ and sufficiently large $\lambda_{\rm CSL}$, indicating that the modified steady-state cannot be accounted for by a reduced quality and other sources of decoherence need to be invoked.
%
%
To quantify these differences we use the distance between the distributions of the mechanical occupations, $D[\bar\mu_{\rm m},\bar\rho_{\rm m}']=\sqrt{\sum_{n=0}^\infty\langle n|(\bar\mu_{\rm m}-\bar\rho_{\rm m}')|n\rangle^2}$, which is plotted as a function of $\lambda_{\rm CSL}$ for three different measurement precisions $\sigma = 10^{-6},~10^{-8},~\text{and}~10^{-10}$ in Fig.~\ref{fig:results}(b).

Of course, the above discussion assumes that the mechanical damping is of a specific form. Yet due to the ample tuneability of our setup, other forms of mechanical damping may also be identified and distinguished from CSL as long as they are not exactly in the form of a momentum diffusion with a rate that is independent of the potential.

%
%
%

%
%
\textit{State verification.---}
As the verification of the prepared states will be of crucial importance in experiments, we now introduce a strategy that allows to determine all elements $P_{mn}$ of steady-state mechanical density matrix $\bar{\rho}_\mathrm{m}=\sum_{mn}P_{mn}\ket{m}\bra{n}$, for which $x_{mn} \ne 0$, and the diagonal elements $P_{nn}$, see \cite{supmat} for the details. We exploit the fact that our proposed architecture is naturally suited to strongly couple the electromagnetic cavity to a superconducting qubit with ground and excited states $|g\rangle$ and $|e\rangle$, and a transition frequency that can be tuned in situ and in real time (e.g., via a time dependent external magnetic flux for transmon and phase qubits \cite{Majer2007,Hofheinz2008}). Measurements of the qubit's population will allow us to read out the mechanical density matrix as follows.

Using an electromagnetic mode that is far detuned from the qubit and mechanical transitions (dispersive regime), one obtains an effective qubit-mechanical interaction of the form $\hat x\sighat_x$, where $\sighat_x=|g\rangle\langle e|+|e\rangle\langle g|$. We show in \cite{supmat} that by initializing the qubit in the ground state, and tuning its frequency to the $|m\rangle\leftrightarrow |n\rangle$ transition ($m>n$), the probability of finding it in the excited state oscillates in time with an amplitude proportional to the diagonal element $P_{mm}$ of the mechanical density matrix. On the other hand, by initializing the qubit in the superposition $|g\rangle+e^{i\varphi}|e\rangle$, the excitation probability becomes sensitive to $i(P_{mn}e^{i(\phi_{mn}-\varphi)}-\text{c.c.})$, where $\phi_{mn}$ is the phase of $x_{mn}$, hence allowing for the determination of the real and imaginary parts of the off-diagonal elements $P_{mn}$ via a proper choice of $\varphi$.

%
%
\textit{Conclusion.---}
We have introduced a method for the steady-state preparation of spatial quantum superposition of a macroscopic object. Our proposal is based on cooling a mechanical mode to the ground state of an engineered double-well potential. We have put forward a specific implementation based on current superconducting circuits and electromechanical technology together with a method for verifying the prepared mechanical state and discussed a strategy for testing the validity of the CSL model. The methods and specific proposal introduced in this work pave the way towards the generation of macroscopic spatial superpositions with available modern technologies, that allow to put bounds on collapse models and shed light on the quantum-to-classical transition.

%
%
\textit{Acknowledgements.---}
We thank Simon Rips, Ignacio Wilson-Rae, Gary Steele, Florian Marquardt, Yue Chang, and Andrea Smirne for useful discussions.
MA and CN-B acknowledge support by the Alexander von Humboldt Foundation via their fellowship for Postdoctoral Researchers. PD-S and MJH have been supported by the German Research Foundation (DFG) via the CRC 631 and the grant HA 5593/3-1.

\appendix
\onecolumngrid
\vspace{1cm}
\begin{center}
\textbf{\Large Dissipative optomechanical preparation of macroscopic quantum superposition states: Supplementary Information}
\end{center}

In this supplemental material we explain details of the concepts discussed in the main text.

\section{Dissipator}
Due to the high degree of anharmonicity present in the double-well potential, a standard term $\diss_{\hat b}$ in the master equation cannot accurately describe the mechanical dissipation in our system. Here we sketch a derivation of the correct terms, in particular obtaining the ones introduced in Eq. (2) of the main text. 
First, we identify the nonlinear mechanical resonator as the system of interest with its local unperturbed dynamics generated by the Hamiltonian $\hat{H}_{\rm dw}$ in Eq.~(1) of the main text.
We model its environment by a set of harmonic oscillators with frequencies $\omega_k$ and annihilation operators $\hat{c}_k$ with the corresponding free Hamiltonian $\hat{H}_E=\sum_k \hbar\omega_k \hatd c_k \hat c_k$. The state of the environment shall be a thermal state at temperature $T$, that is, $\hat{\rho}_E\propto e^{-\hat{H}_E/k_{\rm B} T}$. Further, the system-environment interaction is taken as 
\begin{equation}
\hat H_I/\hbar=\sum_k \hat{x} (g_k \hat{c}_k+g^*_k \hatd{c}_k)= \hat{x}\hat{\mathcal{X}}_E,
\end{equation}
where we have introduced the abbreviation $\hat{\mathcal{X}}_E=\sum_k (g_k \hat{c}_k+g^*_k \hatd{c}_k)$. Note that both $\omega_k$ and $g_k$ are left unspecified at this stage, since only some special combination of them plays a relevant role, and will be chosen later. The full quantum state of the system-reservoir setup shall be represented by $\hat{R}(t)$. In order to eliminate the environmental modes and find an effective master equation for the mechanical state $\hat{\rho}_\text{m}(t)$ we proceed as follows. We first define the Mori projector $\mathcal{P}[\cdot]=\hat{\rho}_E\otimes \text{Tr}_E\{\cdot\}$ whose action on the full state $\hat{R}(t)$ gives $\mathcal{P}[\hat{R}(t)]=\hat{\rho}_E\otimes\hat{\rho}_\text{m}(t)$. Applying this projector and its complement $1-\mathcal{P}$ onto the Liouville equation describing the full dynamics of both system and reservoir, and formally integrating the $1-\mathcal{P}$ projection, we obtain an exact equation of motion for $\hat{\rho}_\text{m}(t)$, the so-called Nakajima-Zwanzig equation \cite{Breuer2007}. The Nakajima-Zwanzig equation features a Dyson series which can be expanded in powers of the system-reservoir interaction $\hat H_I$, and following the standard procedure we apply a Born approximation which takes into account terms up to second order in the interaction. The resulting equation reads
\begin{equation}\label{NZE2}
\partial_t\hat{\rho}_\text{m}(t)=\left[\frac{\hat{H}_{\rm dw}}{i\hbar},\hat{\rho}_\text{m}(t)\right]-\left[\hat{x},\int_0^t d\tau \, \left(\hat{x}(\tau) \tilde{\rho}_\text{m}(t,\tau) \,\text{Tr}_E\{\hat{\mathcal{X}}_E \hat{\mathcal{X}}_E(\tau) \hat{\rho}_E\}-\text{H.c.}\right)\right],
\end{equation}
where we have defined the interaction-picture operators $\hat{x}(\tau)=e^{\hat{H}_{\rm dw}\tau/i\hbar} \hat x e^{-\hat{H}_{\rm dw}\tau/i\hbar}$, $\hat{\mathcal{X}}_E(\tau)=e^{\hat{H}_E \tau/i\hbar} \hat{\mathcal{X}}_E e^{-\hat{H}_E\tau/i\hbar}$, and $\tilde{\rho}_\text{m}(t,\tau)=e^{\hat{H}_{\rm dw}\tau/i\hbar}\hat{\rho}_\text{m}(t-\tau)e^{-\hat{H}_{\rm dw} \tau/i\hbar}$. Note that the first-order term vanished since it is proportional to $\text{Tr}_E\{\hat{\mathcal{X}}_E\rho_E\}=0$. We proceed by evaluating the environmental correlation function which results in
\begin{equation}\label{corr func}
\text{Tr}_E\{\hat{\mathcal{X}}_E \hat{\mathcal{X}}_E(\tau) \hat{\rho}_E\}=\sum_k \left[e^{-i \omega_k \tau} |g_k|^2 (\Nbos(\omega_k)+1)+e^{i\omega_k \tau} |g_k|^2 \Nbos(\omega_k)\right].
\end{equation}
with $\bar{N}(\Omega)=[\exp(\hbar\Omega/k_{\rm B}T)-1]^{-1}$. Next we examine $\hat{x}(\tau)$, for which we introduce the eigenstates $\ket{n}$ and eigenvalues $E_n$ of the double-well potential defined by $\hat{H}_{\rm dw}\ket{n}=E_n\ket{n}$ for $n\in\{0,1,2,...\}$. In practice, we obtain these states numerically, truncating at a certain number of states which are required in order for our simulations to converge. We then obtain
\begin{align}\label{int pic x0}
\hat{x}(\tau) &=\sum_{m>n}\left(x_{mn} e^{-i\delta_{mn}\tau} \ket m\bra n + \text{H.c.}\right),
\end{align}
where $x_{mn}=\bra m \hat{x}\ket n$ and we have ordered the sum over the indices $n$ and $m$ such that the level spacings $\delta_{mn}=(E_m-E_n)/\hbar$ are positive. The terms with $n=m$ vanish as $\bra n \hat{x} \ket n=0 \hspace{2mm}\forall n$ due to the non-degeneracy of $\hat{H}_{\rm dw}$ and its symmetry under the parity transformation $\hat x\to -\hat x$.

The final steps towards obtaining the correct form of the mechanical dissipator are as follows. We insert Eqs.~(\ref{corr func}) and (\ref{int pic x0}) into Eq.~(\ref{NZE2}), and neglect all terms rotating with $e^{\pm i(\delta_{mn}+\omega_k)\tau}$ (rotating-wave approximation) under the assumption that the system-environment interaction is small, $|x_{mn} g_k|\ll \delta_{mn}+\omega_k$. All the remaining terms in Eq.~(\ref{NZE2}) show a dependence on the environmental memory function of the type $f_{mn}(\tau)=\sum_k |g_k|^2 e^{\pm i(\delta_{mn}-\omega_k)\tau} (\Nbos(\omega_k)+1)$. The memory function $f_{mn}(\tau)$ is assumed to decay at a much faster rate than the rate of perturbation due to the system-reservoir interaction for every transition $\ket m\leftrightarrow\ket n$. Apart from validating the Born approximation (which neglects any back-action onto the environment that relaxes back to its unperturbed state $\hat{\rho}_E$ on a much shorter time scale than the one it needs to react to the perturbation $\hat H_I$), it allows us to introduce the Markov approximation \cite{Breuer2007} and approximate $\tilde\rho_\text{m}(t,\tau)\approx \hat{\rho}_\text{m}(t)$ within the environmental memory time. In addition, we perform the time integral in Eq.~(\ref{NZE2}) by bringing its upper limit to infinity, since we are interested in the long time term dynamics, and neglecting any imaginary contributions (Lamb shifts) coming from its principal value.

The validity of the Born-Markov approximation relies on the properties of the microscopic theory at hand, in particular on the interplay between the frequency differences $\delta_{mn}-\omega_k$ and the spectral density $J(\Omega)$, defined by the continuum limit $\sum_k |g_k|^2\to\int d\Omega J(\Omega)$. In the framework of this paper, we focus on microscopic physical scenarios with an Ohmic spectral density $J(\Omega)\propto\Omega$ and where the Born-Markov approximation is very well justified for oscillation frequencies on the order of the harmonic motion around each well at frequency $\omega_0$. In the parameter regimes of interest the level spacings $\delta_{mn}$ of the nonlinear oscillator are comparable to this frequency $\omega_0$ for quantum numbers $n$ and $m$ close to each other. In any other case the resulting terms can be neglected because the overlap $\bra m \hat{x}\ket n$ rapidly tends to zero as $|m - n|$ increases. Overall, we expect the Born-Markov approximation to be well justified for our setup.

Following the steps explained above, we turn the integro-differential master equation (\ref{NZE2}) into the following Markovian master equation with a dissipator for every transition $\ket m\rightarrow\ket n$, reading
\begin{equation}\label{effME}
\partial_t\hat{\rho}_\text{m}=\left[\frac{\hat{H}_{\rm dw}}{i\hbar},\varrho\right]+\frac{1}{2} \Bigg[\hat{x},\,\hat{\rho}_\text{m}\,\underbrace{ \left(\sum_{m>n} \tilde\gamma_{mn} x_{mn}
\left[\bar{N}(\delta_{mn}) \ket m\bra n+(\bar{N}(\delta_{mn})+1) \ket n\bra m\right]\right)}_{=\hat{A}}-\hat{\rho}_\text{m} \hatd{A}\Bigg],
\end{equation}
where we have introduced dissipation rates $\tilde\gamma_{mn} = 2\pi J(\delta_{mn})=(2m\omega/\hbar)\gamma_{mn}$ for the `independent' transitions of frequency $\delta_{mn}$, which contain all the physical constants of the microscopic theory. By adding the terms corresponding to the electromagnetic modes optomechanically coupled to the mechanical system, we finally arrive to Eq.~(2) of the main text.

The dissipator introduced in Eq.~(\ref{effME}) predicts reasonable results as compared to the case of the standard dissipator. In particular the steady-state of Eq.~(\ref{effME}) will be a thermal state with respect to $\hat{H}_{\rm dw}$, i.e. $\lim_{t\to\infty}\hat{\rho}_\text{m}(t)= e^{-\hat{H}_{\rm dw}/k_{\rm B} T}/\tr\big\{e^{-\hat{H}_{\rm dw}/k_{\rm B} T}\big\}$, as we have checked numerically. For $T=0$ it is straightforward to verify that the ground state $\ket 0 \bra 0$ of the nonlinear resonator is indeed the steady-state of the dynamics generated by Eq.~(\ref{effME}). Thus, subject to a zero-temperature bath the nonlinear resonator indeed relaxes to its quantum ground state. Such a result would not have been predicted by the standard dissipator $\mathcal{D}_{\hat{b}}$.

\section{Cooling}

In the main text we commented on the fact that while in principle it should be possible to cool down the mechanical motion to the ground state of the double-well simply by performing sideband cooling on the lowest mechanical transition (the thermal environment would take care of the rest of the transitions in order for the system not to experience effective negative temperatures), this would in practice require exceedingly efficient sideband cooling. In particular, we pointed out that due to the small energy difference between the ground and first-excited states, it is hard to get sufficiently deep into the resolved sideband regime. In this section we elaborate on this point.

Let us consider one electromagnetic mode of a cavity driven off-resonance with a detuning $\Delta$, coupled to the mechanical element through a Hamiltonian of the type $\hbar g\hat x(\hat a +\hatd a)$, where $\hat a$ is the annihilation operator of the cavity mode. Sideband cooling is readily understood within the master equation formalism by adiabatically eliminating the optical mode, under the assumption that its decay rate $\kappa$ is the dominant incoherent rate of the problem. Such an approach would lead to an effective mechanical master equation with both incoherent cooling and heating processes contributing to the different transitions $\ket{m}\rightarrow\ket{n}$ with rates \cite{Rips2014}

\begin{equation}
	\Gamma^{\mp}_{mn} = \frac{g^2|x_{mn}|^2\kappa}{4(\Delta \pm\delta_{mn})^2 +\kappa^2},
\end{equation}
respectively. The ratio between the cooling and heating rates is then
\begin{equation}
	\frac{\Gamma^{-}_{mn}}{\Gamma^{+}_{mn}} = \frac{4(\Delta-\delta_{mn})^2 +\kappa^2}{4(\Delta+\delta_{mn})^2 +\kappa^2}\equiv r_{mn},
\end{equation}
which is the quantity determining how good the cooling process is, that is, how much population is left in the $\ket{m}$ state (smaller the larger this ratio is). We see that optimal cooling is obtained by choosing a detuning $\Delta=-\delta_{mn}$, but even under such conditions it will be effective only provided that $4\delta_{mn}\gg\kappa$.

\begin{figure}[t]
\includegraphics[width=0.6\columnwidth]{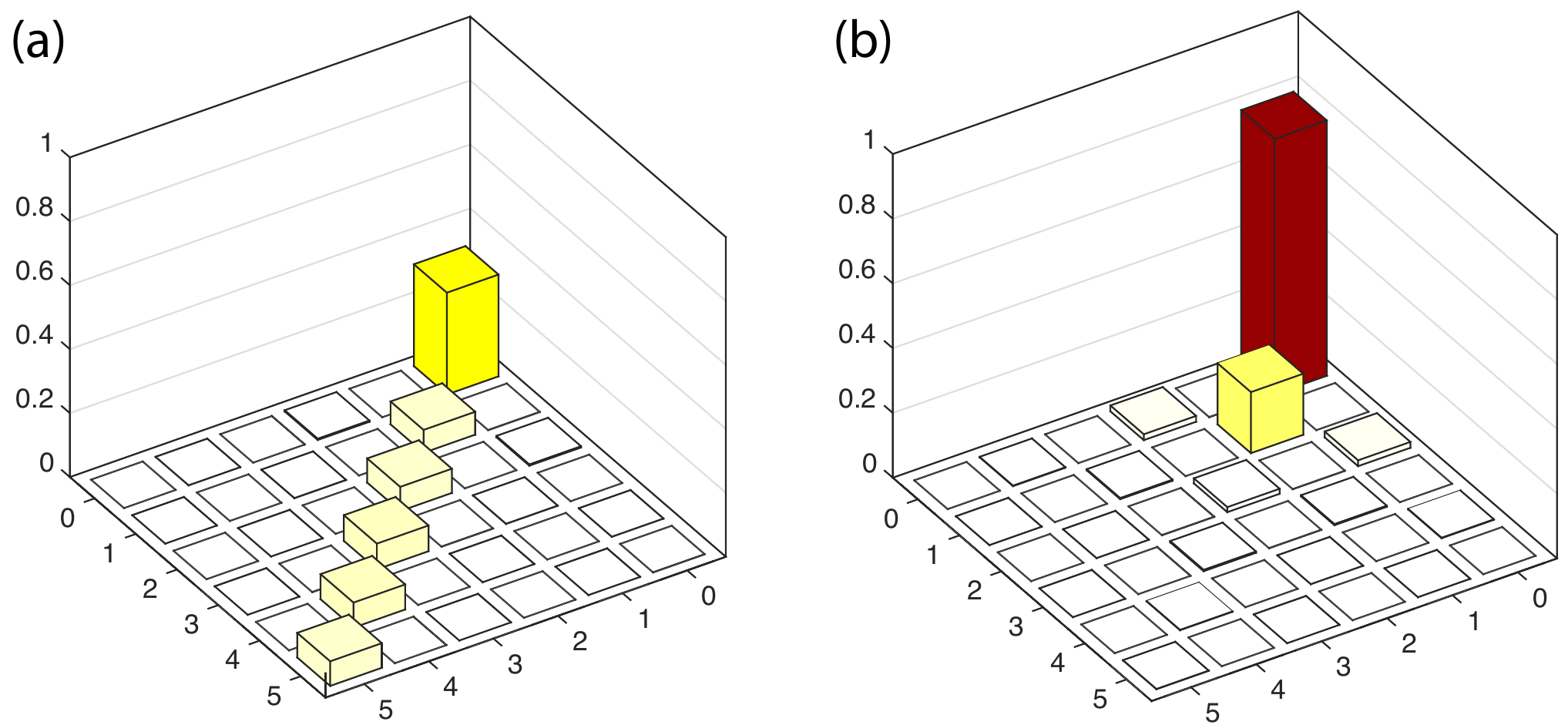}
\caption{Absolute value of the lowest elements of the mechanical steady-state density matrix when cooled by (a) a single electromagnetic mode cooling the lowest transition with optimal coupling rate, or (b) three electromagnetic modes with optimal detunings and optomechanical couplings. See the text for the parameters.}
\label{fig:DensityMatrices}
\end{figure}

As a specific example, let us consider the parameters introduced in the main text for the double-well potential ($m\approx 5.7\times 10^{-16}$~gr, $\nu\approx 1.3\times 10^{-5}$~J/m$^2$, and $\beta\approx 3.7\times 10^{15}$~J/m$^4$), and a cavity linewidth $\kappa=0.3\delta_{10}$. We obtain the ratios $r_{10}\approx 179$, $r_{21}\approx 3092$, $r_{32}\approx 3012$, and $r_{30}\approx 15329$. One can appreciate that the ratio for the lowest transition is sensibly smaller than the others. However, one may think that it could still be large enough. Whereas for the equidistant spectrum of a harmonic potential or a spectrum formed by only two levels, this would indeed be the case, the ratio $r_{10}$ is unfortunately insufficient for performing efficient cooling to the ground state of the nonlinear multilevel spectrum of our device as we have tested by numerically finding the steady-state of the optomechanical master equation with a single cavity mode,
\begin{equation}
	\partial_t\hat{\rho} = \frac{1}{i\hbar}\left[\hat H_{\rm dw}-\hbar\Delta\hatd a\hat a +\hbar g(\hat a +\hatd a)\hat x,\hat{\rho}\right] +\frac{\kappa}{2}\mathcal D_{\hat a}[\hat{\rho}] +\frac{1}{2}\liov_{\rm m}[\hat{\rho}],
\label{master1}
\end{equation}
where $\mathcal L_\text{m}$ is the mechanical dissipator derived above.
In Fig.~\ref{fig:DensityMatrices}(a) we show the absolute value of the elements of the steady-state density matrix using the parameters specified above, addressing the lowest transition ($\Delta=-\delta_{10}$), and optimizing the coupling $g$. Clearly, the ground and first-excited state populations are very similar, showing that cooling is very inefficient in spite of the apparently large $r_{10}$ ratio.

This result can be compared with the one obtained when using three cavity modes with the same linewidth $\kappa=0.3\delta_{10}$, shown in Fig. \ref{fig:DensityMatrices}(b). In particular, we have found that cooling to the ground state is optimized when the cavity modes address the $\ket{1}\rightarrow\ket{0}$, $\ket{2}\rightarrow\ket{1}$, and $\ket{3}\rightarrow\ket{0}$ transitions, with the couplings specified in the main text, where cooling to the ground state is much more efficient than in the previous case.

\section{Elasticity}
Let us now present the mechanical model describing the motion of a thin layer of the type we use in the main text. Denoting by $\zeta(x,y)$ the field describing the vertical displacement of point $(x,y)$ of the layer with respect to its resting plane, its deflection obeys the equation of motion \cite{Landau1975},
\begin{equation}
	\rhotd \partial_t^2 \zeta = -D \nabla^2\nabla^2\zeta +T\nabla^2\zeta,
\label{memeqmot}
\end{equation}
where $\nabla^2$ is the two-dimensional (2D) Laplacian, $\rhotd$ is the membrane's 2D mass density, and $D = Y h^3/[12(1-\sigma^2)]$ with $Y$ being the Young modulus, $h$ the thickness of the membrane, and $\sigma$ its Poisson ratio. Here, $T = T_0 +\Delta T$ is the total tensile force experienced by the membrane, with $T_0$ the force applied by the support and $\Delta T$ the bending tension as a result of extension. The latter is proportional to the relative change in the surface area of the membrane,
\begin{equation}
	\Delta T = Yh\frac{\Delta A}{A_0} = \frac{Y h}{2A_0}\int_0^{2\pi}\hspace{-2mm}d\theta\int_0^a \hspace{-2mm}rdr\left|\nabla\zeta\right|^2.
\label{bendten}
\end{equation}

\subsection{Dominant built-in tensile force}
As it is the practical case, we first assume that the bending energy is much smaller than the contribution coming from the tensile force on the edges of the membrane ($D\nabla^2\nabla^2\zeta \ll  T_0 \nabla^2\zeta$). Thus the equation (\ref{memeqmot}) is simplified to 
\begin{equation}
	\rhotd \partial_t^2 \zeta = T_0\nabla^2\zeta +\left(\frac{Y h}{2A_0}\int_0^{2\pi}\hspace{-2mm}d\theta\int_0^a \hspace{-2mm}rdr\left|\nabla\zeta\right|^2\right)\nabla^2\zeta,
\label{effmemeqmot}
\end{equation}
and leads to nonlinearities in the equation of motion, which can be treated perturbatively in our regime of interest. In particular, we proceed by finding the normal modes associated to the linear problem $\rhotd \partial_t^2 \zeta = T_0 \nabla^2\zeta$, expand equation (\ref{effmemeqmot}) in these, and keep leading nonlinear terms only. 

The membrane geometry is taken to be a disc with radius $a$, which possess radially-symmetric (axisymmetric) modes as well as excited modes with trigonometric polar-angle dependence. The fundamental mode is axisymmetric. We focus on this mode for simplicity, and for practical reasons: (i) its small oscillation frequency leads to large zero-point fluctuations as required in our proposal; (ii) the frequency differences between this mode and the higher modes are large, leading to negligible inter-mode couplings, as discussed in the next section (this feature is also relevant for measurement purposes); (iii) finally, because of its symmetry, it is easier to manipulate this mode in the setup considered here. 

\begin{figure}[b]
\includegraphics[width=0.8\columnwidth]{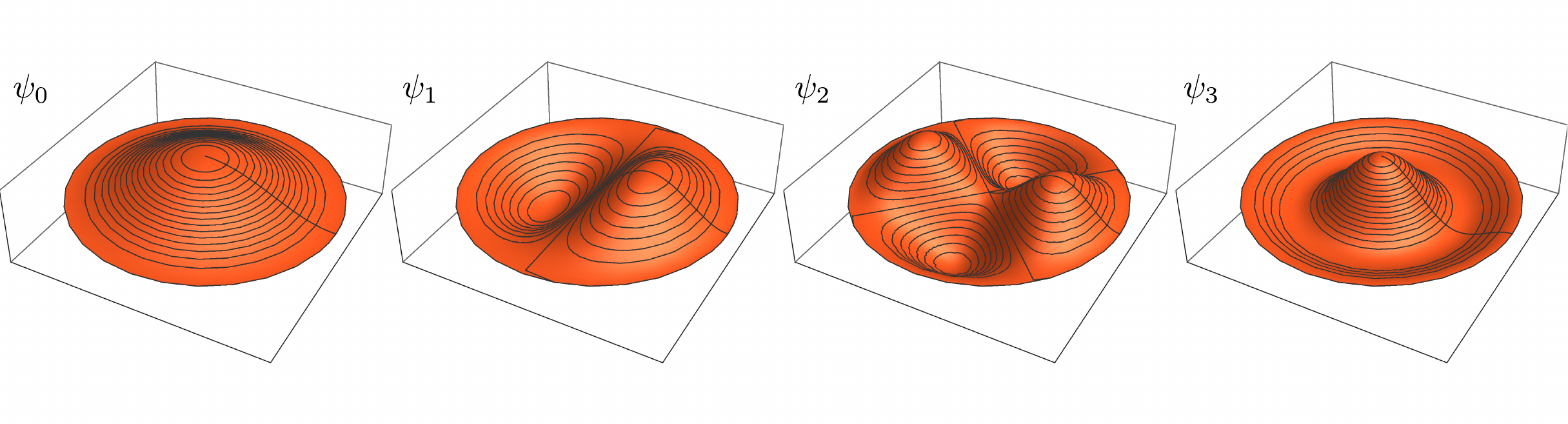}
\caption{Four lowest normal modes of a clamped circular membrane.}
\label{fig:modes}
\end{figure}

We apply the separation of variables $\zeta(r,\theta,t) = \sum_{n=0}^\infty X_n(t)\psi_n(r,\theta)$ to obtain the following normal-mode equations for the linear part of Eq.~(\ref{effmemeqmot}),
\begin{equation}
	\nabla^2\psi_n(r,\theta) +\left(\frac{\lambda_n}{a}\right)^2\psi_n(r,\theta) =0,
\label{modes}
\end{equation}
where we choose to write the separation constant as $\lambda_n^2 = \rhotd a^2\Omega_n^2/T_0$ (to be found numerically as explained below), with a frequency $\Omega_n\in\mathbb{R}$ that will be shown to be the oscillation frequency of mode $n$. The nonsingular solutions of this equation can be written as $\psi_n(r) = \mathcal A_n J_{k_n}(\lambda_n r/a) \cos k_n\theta$, where $J_k$ are Bessel functions of the first kind, $k_n\in\mathbb{Z}$. We order the solutions with increasing values of their separation constant, so that $\lambda_0 < \lambda_1 < \lambda_2 < \cdots $, i.e. the normal modes $\{\psi_n\}_{n=0,1,2,\dots}$ are ordered from lower to higher frequency. For our clamped circular membrane, the boundary conditions imply
\begin{equation}
	J_{k_n}(\lambda_n) =0.
\end{equation}
We choose $\mathcal A_n$ such that $\max\{\psi_n\}=1$ so that $X_n$ provides the maximal vertical deflection of the membrane when the corresponding mode is excited.
The four lowest normal modes are shown in Fig.~\ref{fig:modes}.

The dynamics of each mode is obtained by inserting $\zeta(r,\theta,t)$ into Eq.~(\ref{effmemeqmot}), multiplying both sides of the resulting equation by $\psi_m(r,\theta)$, and integrating over the surface of the membrane. This leads to
\begin{equation}
	m_n^*\ddot{X}_n = -\mathcal K_{n}X_n -\frac{Y h}{2\pi a^2}\sum_{m} \mathcal M_{mm}\mathcal M_{nn} X_m^2X_n.
\end{equation}
Here $m_n^* = \rhotd \int_0^{2\pi}\hspace{-1mm}d\theta\int_0^a \hspace{-0.8mm}rdr \psi_n^2$ is the effective mass and $\mathcal K_{n} = T_0 \mathcal M_{nn}$ the mode's spring constant, where we have introduced $\mathcal M_{mn} = \int_0^{2\pi}\hspace{-1mm}d\theta\int_0^a \hspace{-0.8mm}rdr\psi_m \nabla^2\psi_n = \int_0^{2\pi}\hspace{-1mm}d\theta\int_0^a \hspace{-0.8mm}rdr(\nabla\psi_m)\cdot(\nabla\psi_n)$. We note that the integral $\mathcal M_{mn}$ vanishes for $m \neq n$ as a result of the orthogonality of the modes, which simplifies the nonlinear cross-coupling between the modes to only two modes (instead of four-wave mixing that would appear for the $T_0=0$ case).
Therefore, the dynamics of the membrane can be described by the Hamiltonian
\begin{equation}
	H_{\rm m} = \sum_{n=0}^\infty\Big(\frac{P_n^2}{2m_n^*} +\frac{1}{2}m_n^*\Omega_n^2 X_n^2 +\frac{1}{4}\beta_n X_n^4 \Big) +\frac{Y h}{2\pi a^2} \sum_{n,m=0}^\infty \mathcal M_{nn}\mathcal M_{mm} X_n^2 X_m^2. \label{MultimodeH1}
\end{equation}
where the terms with $n=m$ are excluded from the second sum as they are already included in the first sum. Moreover, we have defined $\Omega_n=\sqrt{\mathcal{K}_n/m_n^*}$ and $\beta_n=(Y h/2\pi a^2)\mathcal M_{nn}^2$. For the fundamental mode $n=0$, which we have taken as our mechanical mode in the main text, we obtain the following expressions for the Hamiltonian parameters in terms of physical parameters:
\begin{subequations}\label{parameters}
\begin{eqnarray}
m_0^* &\approx & 0.27\times (\pi a^2)\rhotd,\\
\Omega_0 &\approx & 2.4\times\sqrt{\frac{T_0}{\rhotd a^2}},\\
\beta_0 &\approx & 3.8\times \frac{Yh}{a^2}.		
\end{eqnarray}	
\end{subequations}
Note that these three parameters have been denoted by $m$, $\omega$, and $\beta$ in the main text. As we discuss below, the interaction between this mode and the remaining ones is negligible for the relevant system parameters, which is why we can treat it as an isolated mechanical mode in the main text.

One could instead strictly work with the elastic equations governing the motion of the membranes displacement and find the Duffing equation for the deflections \cite{Eriksson2013}. However, the results obtained by our simple model here are very similar to the numerical results obtained in Ref.~\cite{Eriksson2013}. In particular, the prefactor in the cubic nonlinear coefficient of the ground state is $\beta_0 \approx 3.9\times Yh/a^2$, which is very similar to what we have computed here. Therefore, we rely on the simple model given in this section for further analysis of the system.

\subsection{Negligible built-in tension}
For the sake of completeness, here we also consider the situation where the circular LDG sheets are fabricated with negligible tensile force $T_0=0$. This leads to different mode profiles and different resonance frequencies. We follow an analogous procedure to the one of the preceding section and first find the normal modes of the linear problem $\rhotd \partial_t^2 \zeta = -D \nabla^2\nabla^2\zeta$ to then expand the equation of motion (\ref{memeqmot}) in these.
We again apply the separation of variables $\zeta(r,\theta,t) = \sum_{n=0}^\infty \tilde X_n(t)\tilde\psi_n(r,\theta)$ to obtain the normal-mode equations,
\begin{equation}
	\nabla^2\nabla^2\tilde\psi_n(r,\theta) - \left(\frac{\tilde\lambda_n}{a}\right)^4\tilde\psi_n(r,\theta) =0,
\label{modes1}
\end{equation}
where now the separation constant is $\tilde\lambda_n^4 = \rhotd a^4\tilde\Omega_n^2/D$. Here $\tilde\Omega_n$ is the oscillation frequency of mode $n$. The nonsingular solutions of this equation can be written as \[ \tilde\psi_n(r) = \left[\tilde{\mathcal{A}}_n J_{k_n}(\tilde\lambda_n r/a) +\tilde{\mathcal{C}}_n I_{k_n}(\tilde\lambda_n r/a)\right ]\cos k_n\theta, \] where $J_k$ are Bessel functions of the first kind, $I_k(y)=i^{-k}J_k(iy)$ are modified Bessel functions, and $k_n\in\mathbb{Z}$. For a clamped circular membrane, the boundary conditions imply
\begin{equation}
	\left[\begin{array}{ll}
		J_{k_n}(\tilde\lambda_n) & I_{k_n}(\tilde\lambda_n) \\
		J_{k_n}^\prime(\tilde\lambda_n) & I_{k_n}^\prime(\tilde\lambda_n)
	\end{array}\right]
	\left(\begin{array}{l}
		\tilde{\mathcal{A}}_n \\ \tilde{\mathcal{C}}_n
	\end{array}\right) =
	\left(\begin{array}{l}
		0 \\ 0
	\end{array}\right),
\end{equation}
with the prime denoting derivative with respect to the argument. These equations have nontrivial solution only when the matrix has a null determinant. Using the recursion relations for Bessel functions this condition can be written as
\begin{equation}
	J_{k_n}(\tilde\lambda_n)I_{k_n+1}(\tilde\lambda_n) +I_{k_n}(\tilde\lambda_n)J_{k_n+1}(\tilde\lambda_n) = 0,
\end{equation}
which can be numerically solved to obtain the roots $\tilde\lambda_n$, and from them the normal mode frequencies $\tilde\Omega_n$. The normal modes, on the other hand, are 
\begin{equation}
	\tilde\psi_n(r) = \tilde{\mathcal{A}}_n\left\{J_{k_n}\left(\tilde\lambda_n\frac{r}{a}\right) -\frac{J_{k_n}\left(\tilde\lambda_n\right)}{I_{k_n}\left(\tilde\lambda_n\right)} I_{k_n}\left(\tilde\lambda_n\frac{r}{a}\right)\right\}\cos k_n\theta,
\end{equation}
where we choose $\tilde{\mathcal{A}}_n$ such that $\max\{\tilde\psi_n\}=1$ so that $\tilde X_n$ provides the maximal vertical deflection of the membrane when the corresponding mode is excited.

The dynamics of each mode is obtained by inserting $\zeta(r,\theta,t)$ into Eq.~(\ref{memeqmot}), multiplying both sides of the resulting equation by $\tilde\psi_m(r,\theta)$, and integrating over the surface of the membrane. This leads to
\begin{equation}
	\tilde m_n^*\ddot{\tilde X}_n = -\tilde{\mathcal{K}}_{n}\tilde X_n -\frac{Y h}{2\pi a^2}\sum_{mlk} \tilde{\mathcal{M}}_{ml}\tilde{\mathcal{M}}_{nk} \tilde X_m \tilde X_l\tilde X_k,
\end{equation}
where $\tilde m_n^* = \rhotd \int_0^{2\pi}\hspace{-1mm}d\theta\int_0^a \hspace{-0.8mm}rdr \tilde\psi_n^2$ is the effective mass, $\tilde{\mathcal{K}}_{n} = D\int_0^{2\pi}\hspace{-1mm}d\theta\int_0^a \hspace{-0.8mm}rdr(\nabla^2\tilde\psi_n)^2$ is the mode spring constant, and $\tilde{\mathcal{M}}_{mn} = \int_0^{2\pi}\hspace{-1mm}d\theta\int_0^a \hspace{-0.8mm}rdr\tilde\psi_m \nabla^2\tilde\psi_n$ provides the nonlinear inter-mode coupling. We note that the integral $\int_0^{2\pi}\hspace{-1mm}d\theta\int_0^a \hspace{-0.8mm}rdr (\nabla^2\tilde\psi_n)(\nabla^2\tilde\psi_m)$ vanishes for $m \neq n$ as a result of the orthogonality of the modes.
Therefore, the dynamics of the membrane can be described by the Hamiltonian
\begin{equation}
	H_{\rm m} = \sum_{n=0}^\infty\Big(\frac{\tilde P_n^2}{2\tilde m_n^*} +\frac{1}{2}\tilde m_n^*\Omega_n^2 \tilde X_n^2 +\frac{1}{4}\tilde \beta_n \tilde X_n^4 \Big) +\frac{Y h}{2\pi a^2} \sum_{n,m,k,l=0}^\infty \tilde{\mathcal{M}}_{nk}\tilde{\mathcal{M}}_{ml} \tilde X_n \tilde X_m \tilde X_l \tilde X_k. \label{MultimodeH}
\end{equation}
where the terms with $n=m=k=l$ are excluded from the second sum as they are already included in the first sum. Moreover, we have defined $\tilde\Omega_n=\sqrt{\tilde{\mathcal{K}}_n/\tilde m_n^*}$ and $\tilde\beta_n=(Y h/2\pi a^2)\tilde{\mathcal{M}}_{nn}\tilde{\mathcal{M}}_{nn}$. 

\section{Electrostatic softening}
As explained in the main text, we induce a double-well potential on the fundamental mode of the membrane by adding a softening force which changes the sign of the corresponding quadratic term in the Hamiltonian (\ref{MultimodeH}). Here we show that a practical way of producing such a softening force is by applying an electrostatic field via a tip electrode located above the center of the membrane. Because of its electrical properties, the membrane will then experience an electrostatic potential with the required negative quadratic term. Such an electrostatic field can be created by two electrodes around the membrane, although in practice it is enough to use a single antenna similar to what has been used in \cite{Steele2009}, whose image in the disk electrode beneath the membrane provides the second electrode (see Fig.~2). In the following we present our simple model for this situation.

The antenna can be treated as a very small conducting sphere. The existence of the disk electrode at $z=0$ allows us to find the electric potential for $z>0$ by the method of images. The disk electrode acts like a mirror and one must consider an imaginary sphere with opposite potential sign on the other side of the disk. The resulting potential is symmetric with respect to rotations around the axis of the tip electrode and only depends on the radial coordinate $r$ and the coordinate perpendicular to the rest plane of the membrane $z$. If $d$ is the distance between the antenna and the disk electrode, the total potential reads
\begin{equation}
	\mathcal V(r,z) = b\mathcal V_1\Big(\frac{1}{\sqrt{r^2+(z-d)^2}} -\frac{1}{\sqrt{r^2+(z+d)^2}}\Big),
\end{equation}
where $b$ is the radius of curvature of the antenna tip and $\mathcal{V}_1$ is the electric potential applied to the electrode. We assume a distance $z_0$ between the disk and the membrane. Therefore, the membrane roughly experiences a symmetric field about its equilibrium position $z_0$. The electric field can be calculated from the potential as $\mathbf{E}=-\nabla\mathcal V(r,z)$, and then the electrostatic energy felt by the membrane is given by \cite{Jackson1998}
\begin{equation}
	V_\text{es} = -\int_{V_s} d\mathbf{p}(\mathbf{r},z)\cdot\mathbf{E}(\mathbf{r},z) \approx -h \int_{A_s} d^2\mathbf{r}~\sigma[\mathbf{r},z_0+\zeta(\mathbf{r})] E_z[\mathbf{r},z_0+\zeta(\mathbf{r})] = -\epsilon_0 h \int_0^{2\pi}\hspace{-2mm}d\theta\int_0^a \hspace{-2mm}rdr E^2_z[r,z_0+\zeta(r,\theta)],
\end{equation}
where we have exploited the radial symmetry of the problem, $V_s$, $A_s$, and $h$ denote the volume, area, and thickness of the sheet, and we have neglected the variation of the charge density within the thickness of the sheet, so that we can approximate its dipole moment at any point of the surface by $d\mathbf{p}=\mathbf{e}_zh\sigma(\mathbf{r},z)d^2\mathbf{r}$, where $\mathbf{e}_z$ is the unit vector in the $z$ direction and $\sigma(\mathbf{r},z)=\epsilon_0E_z(\mathbf{r},z)$ is the surface charge density, which has the dependence on the applied electric field characteristic for a conductor (or superconductor). Applying a Taylor expansion of the integrand around the membrane's equilibrium position $z=z_0$, using the expansion $\zeta(r,\theta,t) = \sum_{n=0}^\infty X_n(t)\psi_n(r,\theta)$ of the mechanical field, and keeping terms up to the fourth order in $X_n$, we can rewrite this electrostatic energy as
\begin{equation}
V_{\rm es} = \sum_n \alpha_n^{(1)} X_n +\sum_{nm} \alpha_{nm}^{(2)}X_n X_m  +\sum_{nml} \alpha_{nml}^{(3)}X_n X_m X_l  +\sum_{nmlk} \alpha_{nmlk}^{(4)}X_n X_m X_l X_k+ \cdots,
\label{elstham}
\end{equation}
where an irrelevant constant term is ignored and the coefficients take the form
\begin{equation}
\label{eleccoeff}
	\alpha_{n_1\dots n_N}^{(N)} = -\frac{h \epsilon_0}{N!} \int_0^{2\pi}\hspace{-2mm}d\theta\int_0^a \hspace{-2mm}rdr\frac{\partial^N [E_z(r,z=z_0)]^2}{\partial z^N}\psi_{n_1}(r,\theta)\cdots\psi_{n_N}(r,\theta).
\end{equation}
The first term in (\ref{elstham}) shows a displacement in the equilibrium position of the mode, while the higher order terms present weak interactions between the \textit{bare} modes. By diagonalizing the quadratic part of the Hamiltonian of the sheet in the presence of electrostatic fields one can find the new normal mechanical modes. However, for the parameters considered in this work, the induced corrections to the mode shapes are found to be negligible (see below), and the same applies to the nonlinear couplings present in Eq. (\ref{elstham}). We thus focus on the contribution to the fundamental mode, simplifying the notation of the electrostatic coefficients to
$\alpha^{(N)}_{0\dots0} \equiv \alpha_{N}$ for $N=1,2,\dots$
obtaining the electrostatic potential provided in the main text, $\hat{V}_{\rm es} = \sum_{j=1}^\infty\alpha_j\hat x^j$, where the notation $\hat x = \hat X_0$ is adopted here and all over the work. A proper choice of the applied voltage $\mathcal V_1$ is able to provide the required (negative) value of $\alpha_2$, yet negligible $\alpha_{N}$ for $N>2$ (see below), to get the desired double-well potential in Eq.~(1).

Note that the above analysis of the electrostatic forces is independent of the normal modes. That is, in expanding the out of plane deflection in different normal modes $\tilde\psi_{n_j}$ as $\zeta(r,\theta,t) = \sum_{n=0}^\infty \tilde X_n(t)\tilde\psi_n(r,\theta)$, different electrostatic coefficients $\tilde \alpha_{n_1\dots n_N}^{(N)}$ should be computed from Eq.~(\ref{eleccoeff}) by  substituting the mode profiles $\psi_{n_j}$ with the new mode profiles $\tilde\psi_{n_j}$. This allows us to compute the electrostatic coefficients via Eq. (\ref{eleccoeff}) for both cases we consider, non-vanishing and vanishing built-in membrane tension.

\section{Application to state-of-the-art setups}\label{AppParam}
\subsection{Dominant built-in tension}
We apply the expressions in Eqs.~(\ref{parameters}) and (\ref{eleccoeff}) to a $a=1~\mu$m sheet of monolayer lithium decorated graphene (LDG), which has $h\approx0.34$~nm, $Y\approx 0.9$~TPa, $\rhotd \approx \rho_{3\rm D}h \approx 7.63 \times 10^{-7}$~kg/m$^2$, and $\sigma \approx 0.24$ \cite{Kganyago2003,Qi2010}.
From these physical parameters, we obtain the Hamiltonian parameters $m\approx 5.7\times 10^{-16}$~gr, $\omega/2\pi \approx 26$~MHz, and $\beta \approx 5.7\times 10^{15}$~J/m$^4$ when the membrane is under a tensile force with $T_0 \approx Yh \times 10^{-5}$ \cite{Barton2011}. For the electrode we assume $d = 1~\mu$m, so that choosing $b\mathcal V_1\approx 4 \times 10^{-4}$~V$\cdot$m will provide the required values for the electrostatic coefficients $\alpha_j$ (assuming $b=100$~nm, this would require the application of $\mathcal V_1\approx 4$~kV to the antenna, which is reasonable). In particular, we obtain the electrostatic parameters $\alpha_2\approx -1.000134(m\omega^2/2)$, $\alpha_3 \approx -57$~J/m$^3$, and $\alpha_4\approx -2\times 10^{10}$~J/m$^4$. We observe that $|\alpha_4| \ll \beta$, so its effect is negligible on the full mechanical potential. In order to see that $\alpha_3$ is also negligible, just note that its zero-point contribution $|\alpha_3|\xzpm\approx 4\times 10^{-10}$~J/m$^2$ is indeed much smaller than the contributions coming from the second and fourth orders, $\nu/2\approx 7\times 10^{-7}$~J/m$^2$ and $(\beta/4)\xzpm^2\approx 4\times 10^{-8}$~J/m$^2$, where $\xzpm = \sqrt{\hbar/2m\omega_0}$.

To verify the isolation of the fundamental mode from the rest of the modes, let us consider the coupling of the three lowest frequency excited modes $X_1$, $X_2$, and $X_3$ to the fundamental mode $X_0$. There are basically two sources for the inter-mode couplings: the nonlinearity intrinsic to the mechanical problem, see Eq.~(\ref{MultimodeH}), and those induced by the electrostatic fields, see Eq.~(\ref{elstham}). Here, we show that for the parameters considered in this work, such interactions are all negligible. As we have already seen in the previous paragraph for the fundamental mode, the dominant electrostatic contribution is that of $\sum_{nm}\alpha_{nm}^{(2)}X_nX_m$ and the higher order coefficients are negligibly small compared to their intrinsic counterparts. This is also true for other modes and their interactions as we have confirmed numerically. Hence, it suffices to show that the quadratic electrostatic couplings and the quartic intrinsic nonlinear interactions are also negligible. The first can be quantified through the ratio between the size of each Hamiltonian term and the frequency difference between the modes that it connects, that is, $\Lambda_{0m} = 2\alpha_{0m}^{(2)}\mean{X_0X_m}/\hbar(\Omega_m' -\omega_0)$, where $\Omega_n' =\left(\Omega_n^2 +2\alpha_n^{(2)}/m_n^*\right)^{1/2}$ is the softened frequency of mode $n \geq 1$ and $\omega_0$ is the oscillation frequency of the fundamental mode around the wells. Similarly, in the case of the quartic contribution, we can quantify it through
\begin{equation}
\Upsilon_{0m} = \frac{Yh}{2\pi a^2}\abs{\frac{\mathcal M_{00}\mathcal M_{mm} \mean{X_0^2X_m^2}}{\hbar(\Omega_m' -\omega_0)}}.	
\end{equation}
In order to estimate these quantities, we here apply a mean-field approximation $\mean{X_0X_m} \approx \mean{X_0}\mean{X_m}$ and $\mean{X_0^2 X_m^2} \approx \mean{X_0^2}\mean{X_m^2}$. Since the fundamental mode is approximately in a mixture of the ground and excited states of the double well, while the rest of the modes are in thermal equilibrium with the environment (the cavity modes only cool transitions of the fundamental mode), we take $\mean{X_0^k} \approx \xzpm^k$ and $\mean{X_m^k} \approx (k_BT/\hbar\Omega_m')^{k/2}(\hbar/2m_m^*\Omega_m')^{k/2}$. For the parameters specified above, and an environment temperature of $15$mK, we obtain $\{\Lambda_{01}=0,\Lambda_{02}=0,\Lambda_{03}=0.03\}$ and $\{\Upsilon_{01}=8\times 10^{-5},\Upsilon_{02}=1\times 10^{-4},\Upsilon_{03}=5\times 10^{-5}\}$. It thus can be appreciated that the fundamental mode is very well isolated.
For the sake of comparison, we here note that the only coupling which would be harmful is the bilinear coupling of the fundamental mode to the next axisymmetric mode $\psi_3$. Since the two modes are strongly off resonance, the associated effective coupling rate is $(\Lambda_{03})^2(2\alpha_{03}^{(2)}\mean{X_0X_3}/\hbar)$. For the parameters considered here, this results in a $\approx 40$~Hz decay from the fundamental mode. Compared to the thermal decoherence rate $\gamma_{mn}\bar N(\delta_{mn}) \approx 10^3$~Hz, this additional decay rate is indeed negligible.

\subsection{Negligible built-in tension}
To show the versatility of our proposal, we also provide the parameters one would engineer for a membrane without intrinsic tension. Let us assume a LDG sheet with the same radius as considered above. Assuming $T_{0} = 0$, we get $\tilde m\approx 3.9\times 10^{-16}$~gr, $\tilde \omega/2\pi \approx 3.8$~MHz, and $\tilde \beta \approx 3.7\times 10^{15}$~J/m$^4$. To obtain a shallow double-well potential we then engineer the electrostatic field such that $\tilde \alpha_2\approx -1.008(\tilde m\tilde\omega^2/2)$.
These values lead to $\delta_{10} \approx 50$~kHz and $G_0/2\pi \approx 30$~Hz. Now, similarly to the case with built-in tension one employs three cavity modes to cool down the three transitions $\ket{1} \leftrightarrow \ket{0}$, $\ket{3} \leftrightarrow \ket{0}$, and $\ket{2} \leftrightarrow \ket{1}$. Optimized intra-cavity photon numbers of $\bar{n}_{\text{c}1} = 1300$, $\bar{n}_{\text{c}2} = 1800$, and $\bar{n}_{\text{c}3} = 6300$ here lead to a ground state occupation of $P_{00}\approx 0.75$, meaning that the mechanical resonator can be found in the desired spatial-superposition state with $\sim75\%$ probability.

The higher order electrostatic coefficients, i.e., $\tilde\alpha_3 \approx -13$~J/m$^3$ and $\tilde\alpha_4\approx -6\times 10^{7}$~J/m$^4$ are negligible because $|\tilde\alpha_4| \ll \tilde\beta$ and the zero-point contribution of the third order term $|\tilde\alpha_3|\tilde x_{\rm zpm}\approx 8\times 10^{-11}$~J/m$^2$ is much smaller than the contributions coming from the second and fourth orders, $\tilde\nu\approx 2\times 10^{-6}$~J/m$^2$ and $\tilde\beta\tilde x_{\rm zpm}^2\approx 1.7\times 10^{-7}$~J/m$^2$.

Consider the coupling of the three lowest excited modes $\tilde X_1$, $\tilde X_2$, and $\tilde X_3$ to the fundamental mode $\tilde X_0$, the electrically induced quadratic coupling is quantified by $\tilde\Lambda_{0m} = 2\tilde\alpha_{0m}^{(2)}\mean{\tilde X_0\tilde X_m}/\hbar(\tilde\Omega_m' -\tilde\omega_0)$, where $\tilde\Omega_n =\left(\tilde\Omega_n^2 +2\tilde\alpha_n^{(2)}/\tilde m_n^*\right)^{1/2}$ is the softened frequency of mode $n \geq 1$ and $\tilde\omega_0$ is the oscillation frequency of the fundamental mode around the wells. Similarly, the quartic contribution can be quantified by
\begin{equation}
\tilde\Upsilon_{0m} = \frac{Yh}{2\pi a^2}\left\{\abs{\frac{\left(\tilde{\mathcal{M}}_{00}\tilde{\mathcal{M}}_{mm} +2\tilde{\mathcal{M}}_{0m}^2\right)\mean{\tilde X_0^2\tilde X_m^2}}{\hbar(\tilde\Omega_m' -\tilde\omega_0)}} +\abs{\frac{4\tilde{\mathcal{M}}_{00}\tilde{\mathcal{M}}_{0m}\mean{\tilde X_0^3\tilde X_m}}{\hbar(\tilde\Omega_m' -3\tilde\omega_0)}} +\abs{\frac{4\tilde{\mathcal{M}}_{0m}\tilde{\mathcal{M}}_{mm}\mean{\tilde X_0\tilde X_m^3}}{\hbar(3\tilde\Omega_m' -\tilde\omega_0)}}\right\}.	
\end{equation}
In analogy to the cases with built-in tension, we apply a mean-field approximation to the position expectation values. For the parameters specified above and an environment temperature of $15$~mK we obtain $\{\tilde\Lambda_{01}=0,\tilde\Lambda_{02}=0,\tilde\Lambda_{03}=0.036\}$ and $\{\tilde\Upsilon_{01}=0.018,\tilde\Upsilon_{02}=0.005,\tilde\Upsilon_{03}=0.01\}$. One thus observes that here the fundamental mode is also very well isolated.

\section{State verification of the mechanical mode via a superconducting qubit}
In the main text we have explained how our setup is very well suited for the verification of the mechanical state via measurements of a single additional superconducting qubit. In this section we provide further details of this method. We will denote the ground and excited qubit states by $\ket{g}$ and $\ket{e}$, respectively, defining the usual Pauli matrices as $\hat{\sigma}_z=\ket{g}\bra{g}-\ket{e}\bra{e}$, $\hat{\sigma}_x=\ket{g}\bra{e}+\ket{e}\bra{g}$, $\hat{\sigma}^+=\ket{e}\bra{g}$, and $\hat{\sigma}^-=\ket{g}\bra{e}$.

\begin{figure}[t]
\includegraphics[width=0.9\columnwidth]{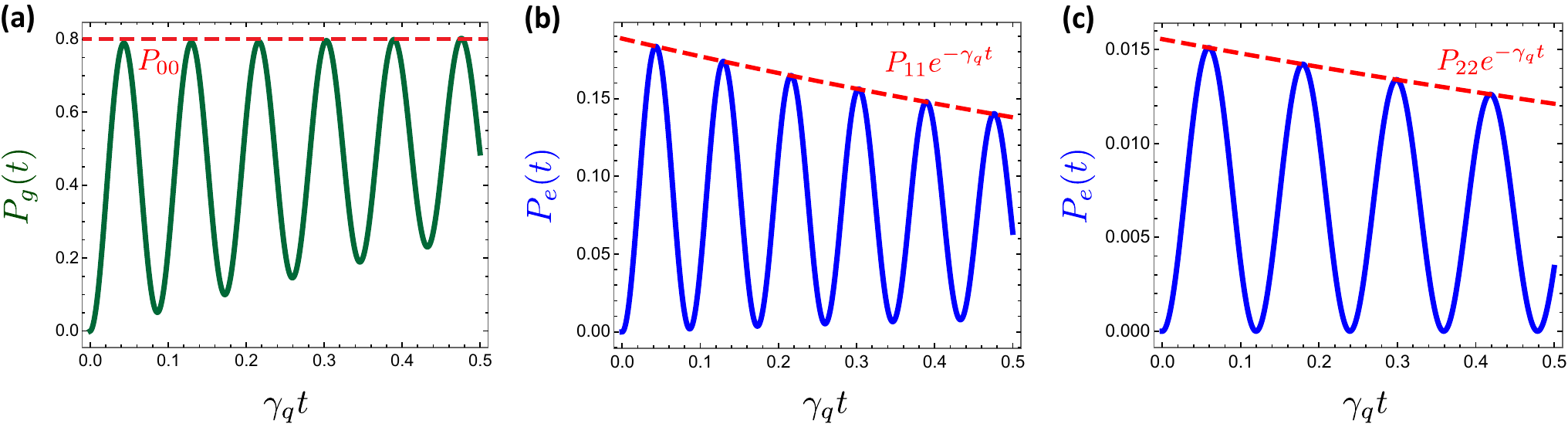}
\caption{The occupation probability of (a) the qubit's ground state ($P_g$), (b) and (c) the qubit's excited state ($P_e$) versus normalized interaction time. In (a) and (b) the mechanical transition $\ket 0 \leftrightarrow \ket 1$ is interacting with the qubit, while in (c) the mechanical transition $\ket 1 \leftrightarrow \ket 2$ is interacting with the qubit. The time traces show sinusoidal oscillations with a period that is proportional to the respective matrix elements of $\hat x$. The amplitude envelope  of the qubit's excited state probability decays like $P_{jj}\exp\{-\gamma_q t\}$.}
\label{fig:ringdown}
\end{figure}

The membrane and the superconducting qubit interact indirectly via their coupling to a cavity mode, a different mode than the ones used for cooling purposes. The full Hamiltonian describing such a system is given by
\begin{equation}
\hat{H} = \hat{H}_{\rm dw} -\hbar\Delta_c \hatd{a}\hat{a} -\frac{\hbar\Delta_q}{2}\sighat_z +\hbar g(\hat a +\hatd a)\hat{x} +\hbar\chi(\hat{a}\sighat^+ +\hatd{a}\sighat^-),
\end{equation}
where $\Delta_c$ and $\Delta_q$ are the detunings of the cavity mode and the qubit with respect to the coherent field driving the cavity. The cavity mode can be adiabatically eliminated when it is far detuned from the mechanical resonator and the qubit, that is, when $|\Delta_c-\Delta_q| \gg \chi$ and $|\Delta_c-\omega_0| \gg g\xzpm$, where we recall that $\omega_0 =\sqrt{\nu/2m}$ is the oscillation frequency of the membrane around the minimum of each well, which is on the order of the lowest mechanical transitions, and $\xzpm$ is the corresponding zero-point fluctuation amplitude. The effective qubit-membrane Hamiltonian can be found via a Schrieffer-Wolff transformation consisting in moving to a picture defined by the unitary $\hat{U} = \exp\left\{\frac{g}{\Delta_c -\omega_0}\hat{x}(\hat{a} -\hatd{a}) +\frac{\chi}{\Delta_c -\Delta_q}(\hat{a}\sighat^+ -\hatd{a}\sighat^-)\right\}$, together with a truncation up to second order in the small parameters $g\xzpm/(\Delta_c -\omega_0)$ and $\chi/(\Delta_c -\Delta_q)$, and a restriction of the dynamics to the subspace with zero photons in the cavity (which is expected to stay unpopulated due to the off-resonant interaction). We obtain
\begin{equation}
\hat{H}_{\rm eff} \approx\tilde{H}_{\rm dw}-\frac{\hbar\tilde{\Delta}_q}{2}\sighat_z+\hbar J\hat{x}\sighat_x,
\end{equation}
where we have defined effective detuning $\tilde\Delta_q = \Delta_q -\dfrac{\chi^2}{2(\Delta_c -\Delta_q)}$, and $\tilde{H}_{\rm dw}$ is defined as $\hat{H}_{\rm dw}$ in Eq.~(1), but with $\nu$ replaced by $\tilde\nu = \nu - \dfrac{4m\omega_0 g^2(\Delta_c -2\omega_0)}{(\Delta_c -\omega_0)^2}$. Since the shift in $\nu$ is however negligible for the domain of interest and hence doesn't affect the conclusions presented above concerning the system parameters. We thus continue our analysis with $\hat H_{\rm dw}$. The strength of the effective interaction between qubit and membrane is
\begin{equation}
J = \frac{g\chi (\Delta_c -\Delta_q -\omega_0)}{(\Delta_c -\Delta_q)(\Delta_c -\omega_0)}.
\end{equation}
Based on this effective interaction, the mechanical state can be verified via the following strategy. Let us first write the Hamiltonian $\hat{H}_{\rm eff}$ in the basis of eigenstates of the double-well Hamiltonian, $\hat{H}_{\rm dw}|n\rangle=E_n|n\rangle$,
\begin{align}
	\hat{H}_{\rm eff} &= \sum_{m=0} E_m \proj{m} -\frac{\hbar\tilde\Delta_q}{2} \sighat_z -\hbar J\sighat_x\sum_{m>n}\left(x_{mn}\trans{m}{n} +x_{mn}^* \trans{n}{m} \right) \nonumber \\
				&\approx \sum_{m=0} E_m \proj{m} -\frac{\hbar\tilde\Delta_q}{2} \sighat_z -\hbar J\sum_{m>n}\left( x_{mn}\sighat^-\trans{m}{n} +x_{mn}^* \sighat^+\trans{n}{m} \right),
\end{align}
where, in the second line, we have applied a rotating wave approximation. In the interaction picture, this Hamiltonian reads
\begin{equation}
	\hat{H}_{\rm eff} = -\sum_{m>n}\hbar J x_{mn}e^{i(\tilde\Delta_q -\delta_{mn})t}\sighat^-\trans{m}{n} +\text{H.c.}
\end{equation}
By tuning the qubit frequency into resonance with a specific mechanical transition, its coupling to the rest of transitions can be neglected, which is a valid approximation if the qubit dephasing rate $\gamma_q$ resolves the double-well transition frequencies, $\gamma_q < |\delta_{mn}-\delta_{jk}|$. Let us assume $\tilde\Delta_q = \delta_{jk}$, then we can approximate the effective Hamiltonian by
\begin{equation} \label{eq:EffHamTomo}
\hat{H}_{\rm eff} \approx \hbar J x_{jk}\sighat^-\trans{j}{k} +\text{H.c.}.
\end{equation}
We use this Hamiltonian for the determination of the elements of the mechanical density matrix. The diagonal elements - except for the ground state - can be found as follows. We prepare the qubit in the ground state, so that the full initial state is $\ket{g}\bra{g}\otimes\bar{\rho}_\text{m}$, where $\bar{\rho}_\text{m}=\sum_{mn} P_{mn}\trans{m}{n}$ is the mechanical steady-state that we want to determine. The element $P_{jj}$ can be determined by looking at the swap dynamics induced by the effective Hamiltonian (\ref{eq:EffHamTomo}). In particular, the probability of finding the qubit in the excited state evolves in time as
\begin{equation}
P_e(t) = P_{jj} \sin^2(\Omega_{jk}t),
\end{equation}
where $\Omega_{jk}=J\abs{x_{jk}}$ is the Rabi frequency and $\ket k$ is a reference state chosen for this measurement via the choice of the detuning $\tilde\Delta_q = \delta_{jk}$. The most convenient choice here would be $k=j-1$ ($j \geq 1$) as it has the maximal matrix element $x_{jk}$ which leads to the largest frequency and hence best resolution in the qubit ring-down curves, see Fig.~\ref{fig:ringdown}.
For measuring $P_{00}$, the occupation number of the mechanical ground state, one instead prepares the qubit in its excited state and observes the swap dynamics. In this case the probability of finding the qubit in its ground state is $P_g(t) = P_{00} \sin^2(\Omega_{0k}t)$, where again $\ket k$ is a reference state and the most convenient choice is $\ket 1$.

To find the off-diagonal elements of the density matrix one needs to prepare the qubit in the superposition state $(\ket{g} +e^{i\varphi}\ket{e})/\sqrt{2}$. In this case, the probability of finding the qubit in the excited state evolves as
\begin{equation}
	P_e(t) = \frac{1}{2}\bigg\{ \sum_{m\neq k}P_{mm} +P_{kk}\cos^2(\Omega_{jk}t) +P_{jj}\sin^2(\Omega_{jk}t) -i\Big[ P_{jk}e^{i(\varphi+\phi_{jk})}-P_{jk}^* e^{-i(\varphi+\phi_{jk})} \Big]\sin(\Omega_{jk}t)\cos(\Omega_{jk}t) \bigg\},
\end{equation}
where $\phi_{jk}$ is the phase associated to the matrix elements of the position operator, $x_{jk}=\abs{x_{jk}}\exp\{i\phi_{jk}\}$. After having found all diagonal elements $P_{jj}$, one can thus determine the imaginary part of $P_{jk}$ by choosing $\varphi =0$ and its real part by choosing $\varphi =\pi/2$. Therefore, two sets of measurements for every off-diagonal element are needed. Note that since $x_{m,m + 2k}=0$ for $k \in \mathbb Z$, not all elements of the density matrix can be measured by this method, preventing it to be a full state tomography.

Let us point out that, in order to observe the Rabi oscillations, one needs a sufficiently large decoherence time of the qubit, as measured by the so-called $T_2^*$ (including both phase and amplitude damping). In particular, $\Omega_{jk} \gtrsim 1/T_2^*$ is the required condition. Here, we assume $\chi /2\pi = 50$~MHz. Since the qubit should be tuned to the mechanical frequencies one has $\tilde\Delta_q \sim \omega_0 \ll \chi$. Therefore, to operate the qubit-cavity interaction in the dispersive regime the cavity must be off-resonance. This will guarantee the dispersivity of the mechanics-cavity interaction provided the electromechanical coupling is small enough $g|x_{mn}|\ll \Delta_c$, which is typically true. For the parameters in our setup, $Jx_{10} \approx 2\pi \times 350$~kHz thanks to the large qubit-cavity coupling $\chi$ we have chosen. This means the dephasing time of the qubit must be $T_2^* \gtrsim 10^{-6}$~s, which is available in the state-of-the-art transmon qubits \cite{Barends2013}.

The feasibility of the state verification scheme can be shown more rigorously by simulating the qubit-mechanical dynamics. The master equation we use for this analysis is
\begin{align}
	\dot\varrho &= \frac{1}{i\hbar}[\hat H_{\rm eff},\varrho] +\frac{1}{2}\liov_{\rm m}[\varrho] +\frac{\gamma_q}{2}\diss_{\hat\sigma^-}[\varrho] +\frac{\tilde\gamma_q}{2} \diss_{\hat\sigma_z}[\varrho],
\end{align}
where $\varrho$ is the bipartite density matrix including the qubit and the mechanical mode and $\tilde\gamma_q$ is the pure dephasing rate of the superconducting qubit. Here, we take $\tilde\gamma_q \approx 2 \gamma_q = 2\pi \times 10$~kHz \cite{Barends2013}. In Fig.~\ref{fig:ringdown} the measurement outcomes for the first three diagonal elements of the the mechanical density matrix are shown. One observes that the state-of-the-art superconducting qubits would allow for resolving the swap oscillations of the mechanical transitions, hence measuring their occupation number.

%
%
\bibliography{buckled}

\end{document}